# Achieving CMMI Level 2 with Enhanced Extreme Programming Approach

Tuomo Kähkönen[1], Pekka Abrahamsson [2]

[1]Nokia Research Center
P.O. Box 407, FIN-00045 NOKIA GROUP, Finland
tuomo.kahkonen@nokia.com
[2]VTT Technical Research Centre of Finland
P.O. Box 1100, FIN-90571 Oulu, FINLAND
Pekka.Abrahamsson@vtt.fi

**Abstract.** The relationship between agile methods and Software Engineering Institute's CMM approach is often debated. Some authors argue that the approaches are compatible, while others have criticized the application of agile methods from the CMM perspective. Only few CMM based assessments have been performed on projects using agile approaches. This paper explores an empirical case where a project using Extreme Programming (XP) based approach was assessed using the CMMI framework. The results provide empirical evidence pointing out that it is possible to achieve maturity level 2 with approach based on XP. Yet, the results confirm that XP, as it is defined, is not sufficient. This study demonstrates that it is possible to use the CMMI for assessing and improving agile processes. However, the analysis reveals that assessing an agile organization requires more interpretations than normally would be the case. It is further concluded that the CMMI model does not always support interpretations in an agile context.

## 1 Introduction

Agile software development approaches have generated a lot of interest in the field of software engineering in the last few years. A number of studies have shown that agile solutions are a viable option for many software companies producing software in a volatile business environment. Volatility has been contrasted with the stability. In fact, one of the more interesting debates in software engineering community is concerned with two apparently very different approaches for software process improvement: CMM[1] promoted by Software Engineering Institute (SEI) [1] and Extreme Programming developed by Beck [2]. CMM is often seen as the arch-type of traditional SW development and contradicted with agile development practices. Many authors have suggested that organizations should develop methods that combine agile and traditional elements [3-10]. Some authors have also argued that in principle the CMM and agile approaches are compatible [3, 8, 10-14]. However, it has proven to be difficult to combine these approaches in practice [3, 10] and many important limitations in the existing agile methodologies, like XP, have been pointed out from the CMM perspective [7, 8, 13-20]. Yet, only few if any studies have performed a CMM-based assessment on an agile project. For this reason currently it is not well understood how to build methods that that combines these two approaches in practice.

---

[1] CMM and Capability Maturity Model are registered in the US Patent and Trademark Office. CMMI and SCAMPI are service marks of Carnegie Mellon University. Term CMM is used in this article to include both Software Capability Maturity Model (SW-CMM) and Capability Maturity Model Integration (CMMI).

The primary purpose of this paper is to increase understanding about the relationship between XP and CMMI. This paper reports results from a study that analyzed using CMMI as the frame of reference a software development project using an enhanced XP based process. The project assessed was rated at a CMMI maturity level 2. The results of this study confirm the theoretical comparisons between XP and CMM [8, 13, 16, 21, 22] claiming that XP does not fulfill CMM requirements. However, the results also show that it is possible to construct a process that fulfills CMMI requirements by adding practices to XP. These additions are outlined in detail so that other organizations can benefit from the results. It is claimed that the results are applicable to certain extent to other agile methods as well. Finally, the challenges of using CMMI for assessing and improving agile processes are discussed.

This paper is organized as follows. Next section presents the research objectives and outlines the objectives of the CMMI. The third section introduces the empirical case project and the fourth section presents the key findings of the study. Section five discusses the implication of these findings and relates them to the literature. The paper in concluded with final remarks.

## 2 Research objectives and methods

This paper aims at increasing understanding about the relationship between XP and CMMI. This is achieved by performing an actual CMMI assessment on a project using an agile method. First research objective is to demonstrate how an enhanced XP based process can satisfy the criteria set by CMMI. This is done comparing the actual practices performed in the project to CMMI requirements practice by practice. The second research objective is outline the challenges of using CMMI for assessing and improving agile processes. This is performed by using the insight and experiences gained from the assessment process.

The research approach used in this study is different from other similar studies [8, 13, 16, 21, 22] in two important aspects: First, the evaluation of the CMMI requirements is done within the context of a real life project. Second, the analysis is taken down to the more detailed practice level from the goal level. This more detailed approach is expected to deliver more reliable results.

A real life project assessment was selected over desk exercise because CMMI highlights that when it is used to enhance existing processes, professional judgment must be used to interpret the CMMI practices. The practices must be interpreted using an in-depth knowledge of CMMI, the discipline, the organization, the business environment, and the specific circumstances involved. [1, 23] Thus it is not possible to obtain universally applicable results portraying that "a method X is or is not CMMI compliant" but the method adequacy and institutionalization must be assessed case by case.

The analysis in this paper covers specific goals and practices of the CMMI maturity level 2 process areas excluding Supplier Agreement Management. The generic goals are not included in the analysis because the research focus is in the method rather than its implementation and institutionalization in the case organization.

CMMI was selected to be the assessment framework firstly because it offers major improvements to SW-CMM regarding iterative and risk driven development[7, 24], and thus it can be supposed to be more aligned with the agile development ideas that SW-CMM. And secondly, because SW-CMM is not developed by the SEI any longer and CMMI is known to replace it in few years time. The next section introduces CMMI briefly.

### 2.1 CMMI

Capability Maturity Models in general contain the essential elements of effective processes for one or more disciplines. These elements are based on the concepts developed by Crosby [25], Deming [26], Juran and [27] Humphrey [28]. CMMI integrates systems engineering, software engineering, and integrated product and process development in one model. The purpose of CMMI is to provide guidance for improving organization's processes and enables the organization to better manage the development, acquisition, and maintenance of products or services. [1]

Maturity level is a central concept in CMM. It is *a priori* defined evolutionary plateau of process improvement. Each maturity level stabilizes a part of the organization's processes. At level 2, i.e. managed, the organization has ensured that its processes are planned, documented performed monitored, and controlled at project level [1].

Each maturity level in the CMMI contains several process areas that have two types of goals: specific and generic. Specific goals apply to one process area and address the unique characteristics that describe what must be implemented in order to satisfy the purpose of the process area. Generic goals apply to all process areas and address the implementation and institutionalization of the process area. [1]

Specific and generic goals are the only required components of the CMMI. In addition there are expected components, specific and generic practices, describing what organization will typically implement to achieve the goals. The actual practices in an organization assessed must be interpreted using an in-depth knowledge of CMMI, the discipline, the organization, the business environment, and the specific circumstances involved. The organization does not need to implement the practices as described in CMMI model. It is acceptable to implement an alternative practice that fulfills the same purpose.[1, 23] The context must be taken in account when evaluating adequacy of proposed alternative practice. The same practice may be adequate in one situation (e.g. small project) and the required goals are fully achieved but in different situation (e.g. large project) the same practice may prove to be inadequate [29].

### 2.2 Research setting

The assessed project was VTT's eXpert project [30]. The assessment was performed after the project had finished, system testing was completed and the software product was in actual use. The scope of assessment was set *a priori* at CMMI level 2. The Supplier Agreement Management key process area (KPA) was excluded, because the project did not have any subcontracting.

Used assessment method (Nokia CMMI-Based Process Assessment, CMMI-B) is based on SCAMPI (Standard CMMI Appraisal Method for Process Improvement) [31] and is supposed to be ARC (Appraisal Requirements for CMMI) Class B compliant [32]. Although the level 2 rating achieved in this assessment is not official SEI rating as only Class A assessment can produce ratings for benchmarking [32], the CMMI-B assessment results are expected to be fairly close to the results of Class A compliant appraisal.

Assessment team included three persons: The first author as the assessment team leader with experience from several CMMI assessments and the second author and a third person as assessment team members. The second author is also trained assessor. The assessment team had available all the material produced by the team including story cards and flip charts. The assessment team familiarized with the material before the interviews and had it available during the rating session.

The assessment team interviewed the project manager, two developers, customer, system test specialist and the business manager. Assessment team members took notes from the interviews to process specific templates. The interviews were also tape recorded (but not transcribed) for research and verification purposes. The interviewees were notified that the results could be published as a research paper.

Ratings were done immediately after assessment and the objective evidence found was written down in an Excel sheet practice by practice.

## 3 Case Description: The eXpert Project

The eXpert project's implementation phase was carried out 3.2.2003 – 11.04.2003 in VTT Electronics Oulu, Finland. The purpose of the project was to develop an intranet application for managing research information based on its logical structure. A team of four developers was acquired from the University of Oulu to implement the project. Table 1 describes the roles of different people/groups involved in the eXpert project. The project had a fixed time contract: 8 weeks in calendar time and 1000 hours effort. The time and effort were thus fixed. The flexibility was reserved for the delivered functionality. The project used an approach suggested by Lippert [33]. XP practices [for more detailed description of the XP practices, see 2, 34, 35]) were extended with some additional practices:

> Before the project started to develop code, a separate planning team had worked with the initiation of the project. This work continued during the project in Steering Group meetings. Steering group had three meetings: at the beginning of the project, in the middle and after the project.

A two-day XP workshop was held for the project team before the start of the project. The project team was walked through the XP practices and the tools to be used with the manager and agreed on the practices to be followed in the project.

One project mission was to collect data from XP process for research purposes, so there were enhanced data collection mechanisms in place. The collected data included time, defect and data [36]

In configuration management (CM) area some additional practices were introduced. These included written CM plan, light CM audit procedure at the end of each iteration and a set of explicit change request/error sheets.

Project had daily wrap-up meetings to discuss progress, plans and problems.

XP supposes that project reflects at times how it is doing and tries to find ways to improve [2]. The eXpert project took this further by organizing a semi-formal post-mortem workshops after every iteration according to the guidelines proposed by Dingsøyr and Hanssen [37]

The project did some additional documentation not typically done in XP projects. The planning team elaborated an implementation plan and the project manager authored a written project plan. Project manager also maintained a spreadsheet called Task Book that contained release plan and planned and actual effort spent for each task. Minutes were taken from the steering group meetings. There was also a CM plan, CM audit checklist, change request log and an error log to help CM activities. Architecture, database and user interface description documents were written during the last iteration and a system test report was made from the system testing. The results of the post-mortem sessions were recorded and displayed on the wall. Later these notes were transformed to a document.

In the end of each iteration the project team had a pre-release testing session. After that the product was released to volunteered end-users for testing and getting feedback. External specialist designed the system testing procedure for the final product. First initial systematized testing was done after iteration 3 and second more comprehensive testing was done at the end of the project.

**Table 1.** Roles in the eXpert project

| Role | Purpose |
|---|---|
| Planning Team | Worked prior to the project; defined projects' scope etc. |
| Steering Group | Members were from the university and the research institute including the project team. |
| Business Manager | Started the project and owned the results. Was responsible for providing all needed facilities and resources for the project. |
| Customer | Key user of the system under construction. Had the best understanding what the system should do. |
| Project Manager | Was team member responsible for project management. |
| CM Specialist | Team member who handled CM issues. |
| Metrics responsible | Team member, ensured that the metrics required were collected. |
| Research responsible | Team member, ensured that the research question the team undertook to solve was addressed. |
| System Test | Planned an executed system tests. |

| Specialist | |
|---|---|
| End User | 17 volunteered end users tested each release and reported found bugs and improvement ideas to the customer. |

## 4 Assessment Results

The eXpert project was rated at CMMI level 2. All specific goals (SG) were fully satisfied although there were minor findings and interpretation issues at specific practice (SP) level. This chapter presents the rationale for the rating of the specific goals practices by practice in the tables 2-14. Each table presents the specific goal on the top, practice definition on the left, and the rationale for rating on the right. The goal and practice descriptions are from [1]. Based on the rating rationale presented in the tables, the reader can verify the accuracy of the proposed CMMI level 2 rating for the project.

**Table 2.** Requirements Management SG1

| SG 1: Requirements are managed and inconsistencies with project plans and work products are identified. ||
|---|---|
| **SP1.1:** Develop an understanding with the requirements providers on the meaning of the requirements | Pre-project planning team defined the project goals and the initial set of the user stories. During the project the customer had the responsibility to provide the requirements. Planning Game was used to communicate the requirements to the team. |
| **SP1.2:** Obtain commitment to the requirements from the project participants | This was achieved in planning meetings. |
| **SP1.3:** Manage changes to the requirements as they evolve during the project | The customer who worked as a part of the team had an active role in managing changes to the requirements. He maintained change request Excel sheet continuously. Changes were communicated to the team in planning game where also the impacts of the change were analyzed. When a story changed, old story card was archived. |
| **SP1.4:** Maintain bidirectional traceability among the requirements and the project plans and work products | Continuously maintained Task Book stated which user stories/tasks are implemented in each release. Releases were uniquely identifiable based on a baseline in the CM system. This enabled bi-directional user story/task – release traceability. There was also manual traceability between user stories and unit tests. |
| **SP1.5:** Identify inconsistencies between the project plans and work | Team and the customer together checked the consistency of the plans and the requirements in the planning game. CM audit checked after each |

| products and the requirements | iteration that all planned/reported user stories and tasks had been implemented. |
|---|---|

Table 3. Requirements Management SG2

| SG2: Estimates of project planning parameters are established and maintained ||
|---|---|
| **SP2.1:** Establish a top-level work breakdown structure (WBS) to estimate the scope of the project | This practice was not applicable because the customer was not able to define requirements exactly in the beginning of the project. The project used an alternative practice. It had fixed effort and variable scope. Planning game was used to define the scope of the work for each iteration. |
| **SP2.2:** Establish and maintain estimates of the attributes of the work products and tasks | Tasks were estimated only for the next iteration. The short estimation cycle with frequent feedback helped to make accurate estimates although estimates based on expert opinions. |
| **SP2.3:** Define the project life-cycle phases upon which to scope the planning effort | Project had incremental fixed release schedule set by planning team. Every iteration formed one phase in the project. |
| **SP2.4:** Estimate the project effort and cost for the work products and tasks based on estimation rationale | The total effort of the project was fixed. How the effort was used was estimated according to XP procedures in planning meetings. As the project had a fixed effort, fixed price contract, the planning variable was the scope of the work. |

Table 4. Project Planning SG1

| SG1: A project plan is established and maintained as the basis for managing the project ||
|---|---|
| **SP1.1:** Establish and maintain the project's budget and schedule | The project had a fixed budged. The schedule was established in planning meetings and documented in Task Book. Tasks were used to determine what can be accomplished within an iteration. |
| **SP1.2:** Identify and analyze project risks | Project manager identified project risks that were documented in the project plan and discussed in steering group. The actions originated from the risks were discussed in post mortem meetings. |
| **SP1.3:** Plan for the management of project data | CM plan identified configuration items and how to manage them. Team agreed practices needed to manage other data (e.g. story cards). |
| **SP1.4:** Plan for necessary resources to perform the project | Planning team did rough estimate of persons needed and the project duration. Project resources were documented in the project plan. Project had an opportunity to use various specialists in VTT as needed. The procedure for this was in project plan. |
| Plan for knowledge and | This planning was done by the planning team and |
| skills needed to perform the project. | documented in the implementation plan. |
| **SP1.5:** Plan the involvement of identified stakeholders | Planning team had planned this in the implementation plan. |
| **SP1.6:** Establish and maintain the overall project plan content | A written project plan existed according to standard VTT project plan guidelines. Project plan was updated after every iteration. Schedules was maintained in Task Book. |

**Table 5.** Project planning SG2

| | |
|---|---|
| **SG2:** Commitments to the project plan are established and maintained. ||
| **SP2.1:** Review all plans that affect the project to understand project commitments | Project was independent from others, so it did not need to review any plan. |
| **SP2.2:** Reconcile the project plan to reflect available and estimated resources | Reconciling was done in planning meetings. The scope of the work was adjusted to mach the resources. |
| **SP2.3:** Obtain commitment from relevant stakeholders responsible for performing and supporting plan execution | Commitments were obtained in steering group review. |

**Table 6.** Project monitoring and control SG1

| | |
|---|---|
| **SG1:** Actual performance and progress of the project are monitored against the project plan. ||
| **SP1.1:** Monitor the actual values of the project planning parameters against the project plan | Actuals (user Stories/tasks/hours) were collected in the Task Book Excel sheet alongside with the estimates. |
| **SP1.2:** Monitor commitments against those identified in the project plan | Project commitments were monitored and adjusted in planning meetings. Problems endangering those commitments were handled in wrap-up meetings. |
| **SP1.3:** Monitor risks against those identified in the project plan | Planned risk mitigation activities were performed. The risks were reviewed in SG meetings. |
| **SP1.4:** Monitor the management of project data against the project plan | CM audits checked the configuration items after every iteration. SG reviewed that documents are done. |
| **SP1.5:** Monitor | Management monitored stakeholder involvement as |
| stakeholder involvement against the project plan | a part of their work. The number of stakeholders was limited and contacts with them frequent. |
| **SP1.6:** Periodically review the project's progress, performance, and issues | Green bar on the wall indicated task progress continuously. Progress and issues were discussed in wrap-up and planning meetings. Every release made the progress visible. Steering group reviewed the progress in their meetings. |
| **SP1.7:** Review the accomplishments and results of the project at selected project milestones | Each release can be considered as a milestone in this case. Review was done in CM audit, post mortem session and in the next planning meeting. |

**Table 7.** Project monitoring and control SG2

| | |
|---|---|
| **SG2:** Corrective actions are managed to closure when the project's performance or results deviate significantly from the plan. ||
| **SP2.1:** Collect and analyze the issues and determine the corrective actions necessary to address the issues | Issues were identified in daily wrap-up meetings and in post mortem sessions. In addition manager visited the project team daily asking for possible issues. Management actions were taken as needed. |
| **SP2.2:** Take corrective action on identified issues | Management actions were done, post mortem sessions resulted actions. |
| **SP2.3:** Manage corrective actions to closure | If the problem was not solved, it was discussed again in wrap-up meeting or post mortem session. Notes from the previous post mortem sessions were on the wall. |

**Table 8.** Measurement and analysis SG1

| | |
|---|---|
| **SG1:** Measurement objectives and activities are aligned with identified information needs and objectives ||
| **SP1.1:** Establish and maintain measurement objectives that are derived from identified information needs and objectives. | The planning team and manager defined initially the measurement objectives. One measurement objective was to collect research data on XP. The measurement objectives were communicated verbally to the project manager. |
| **SP1.2:** Specify measures to address the measurement objectives | Planning team defined an initial set of measures to be collected. The project manager defined some additional metrics that he needed (Task Book). |
| **SP1.3:** Specify how measurement data will be obtained and stored | Project Manager had explicit responsibility to define data collection. Paper-pen method was used to collect the data daily. Some data was obtained from CVS and the Task Book. |
| **SP1.4:** Specify how measurement data will be analyzed and reported | Operative metrics analysis and reporting was built in the Task Book. Research data was analyzed separately from the project. |

**Table 9.** Measurement and analysis SG2

| SG2: Measurement results that address identified information needs and objectives are provided | |
|---|---|
| **SP2.1:** Obtain specified measurement data | Based on the measurement records, the data was obtained very accurately. |
| **SP2.2:** Analyze and interpret measurement data | Operative analysis was done in Excel based on Task Book data. Mostly these were simple graphs. |
| **SP2.3:** Manage and store measurement data, measurement specifications, and analysis results | Operative metrics and analysis were stored on file server. This was because also the manager needed to access it but he did not have access to CVS repository. |
| **SP2.4:** Report results of measurement and analysis activities to all relevant stakeholders | Manager had access to measurement results all the time. Measurement results were presented in post mortem sessions and steering group meetings. |

**Table 10.** Process and product quality assurance SG1

| SG1: Adherence of the performed process and associated work products and services to applicable process descriptions, standards, and procedures is objectively evaluated. | |
|---|---|
| **SP1.1:** Objectively evaluate the designated performed processes against the applicable process descriptions, standards, and procedures | On-site customer followed that team follows the agreed processes. In post mortem session manager and customer evaluated with the team the process and identified possible deviations and improvement needs. Manager followed the metrics in order to identify deviations from the process. |
| **SP1.2:** Objectively evaluate the designated work products and services against the applicable process descriptions, standards, and procedures | Manager and steering group reviewed the documents. They checked e.g. that correct templates have been used and they contain all applicable items (e.g. VTT project plan guidelines). Customer checked using samples in CM audits that agreed coding standards had been used. |

**Table 11.** Process and product quality assurance SG2

| SG2: Noncompliance issues are objectively tracked and communicated, and resolution is ensured | |
|---|---|
| **SP2.1:** Communicate quality issues and ensure resolution of noncompliance issues with the staff and managers | Manager and customer communicated and handled the identified quality issues with the team in post mortem sessions and daily meetings. Manager tracked quality issues. The issue was brought up again if no improvement happened. Quality issues were discussed also in steering group meetings. |
| **SP2.2:** Establish and maintain records of the | There were two types of records: configuration audit reports and post mortem session notes on the |

| quality assurance activities | wall. |
|---|---|

**Table 12.** Configuration management SG1

| SG1: Baselines of identified work products are established. ||
|---|---|
| **SP1.1:** Identify the configuration items, components, and related work products that will be placed under configuration management | Configuration items and their storage were stated explicitly in CM plan. The documents were later stored in file server instead of CVS. A naming convention for different version of the documents was agreed and used. The customer had Change Request Log on his workstation. |
| **SP1.2:** Establish and maintain a configuration management and change management system for controlling work products | Project used CVS to store source code. Change management system was manual. Customer maintained a Change Request Log where all changes and their status were tracked. |
| **SP1.3:** Create or release baselines for internal use and for delivery to the customer | Baselines were created for every release. |

**Table 13.** Configuration management SG2

| SG2: Changes to the work products under configuration management are tracked and controlled. ||
|---|---|
| **SP2.1:** Track change requests for the configuration items | All requirement change management was done through the customer who had a Change Request Log that he used for himself to track the changes. Error log was used to manage error corrections for release candidates/releases. CM audit ensured that agreed changes were actually implemented. |
| **SP2.2:** Control changes to the configuration items | Release baselines in CVS were frozen. Team members were allowed to make a new version of any file or document any time. It was possible to change old versions of documents that located on file server but this risk was acknowledged. |

**Table 14.** Configuration management SG3

| SG3: Integrity of baselines is established and maintained ||
|---|---|
| **SP3.1:** Establish and maintain records describing configuration items | Change Request list, Error list, automatic change history in CVS, manually updated change history in document. |
| **SP3.2:** Perform configuration audits to maintain integrity of the configuration baselines | Customer performed a light but sufficient configuration audits after every release. He had a checklist that was filled as he went through the items. The filled checklist was archived. |

## 5 Discussion

There are two important factors that helped the eXpert project to achieve CMMI maturity level 2. Firstly the project used additional practices that are not part of normal XP process and did some additional documentation. If these practices were not been in use and the documents would not have been there, several goals would not have been rated fully satisfied. Secondly, the size of the project was small and the co-operation with the team members so intense that many light practices were sufficient for the particular situation. If the project had been larger or the communication within the team had been less intense, some of the practices would have been inadequate for the situation.

These results show that although there are several weaknesses in XP from SW-CMM/CMMI perspective, it is possible to achieve CMMI level 2 using a process that is based on XP and is extended with additional practices. Attention should be paid on the generalizability of this result. It can't be concluded that the use of eXpert process automatically lead organization to CMMI level 2 maturity. This is because the adequacy of the practices depends on the context where they are used. For larger projects, or projects operating in otherwise more challenging environments, some additional/alternative practices may be needed.

It was found out that CMMI can be used to assess processes that are combining agile and traditional elements. However, based on the assessor experiences, this is a very challenging task. Although the assessor had experience both from CMMI assessments and XP, interpreting mandatory and expected CMMI elements in the XP context was not always easy or straightforward. Also, because the informative information in CMMI was of little help, the role of assessor interpretations was emphasized.

Many of the interpretation issues in this assessment were culminating on knowledge management. CMMI and agile methods manage knowledge created during the projects differently [8]. CMMI places emphasis on explicit, documented knowledge [7, 24]. Many practices that explicitly create documents are expected and also the definition of a work product in CMMI suggests that the work products must be concrete artifacts files or documents [1].

Research has shown that software development is highly knowledge-intensive work [e.g., 38, 39] and centric role of tacit (i.e. undocumented) knowledge in agile method has been highlighted [7, 40, 41]. The role of tacit knowledge is seen to be very important in knowledge creation process in general [42-44]. The tacit knowledge is manifested in different kinds of intangible artifacts and concrete artifacts like documents present only a tip of the iceberg of the whole knowledge [44].

The question the assessor has to answer is whether an alternative practice, that relies on tacit knowledge, and can be considered working and institutionalized, is acceptable or not. Currently when there are no common guidelines for interpreting adequacy of agile practices, different assessors may end up with different ratings.

Although professional judgment is an integral part of CMMI assessment, Turner [8] has suggested that there are two different CMMI schools regarding agility – a conservative, by-the-letter group and a liberal, concept-oriented group. This makes ratings more assessor-dependent thus hampering the reliability of the assessment results. This limitation is mitigated in this study by reporting the rationale for the rating level achieved. To our believe, the CMMI vs. agile discussion will benefit from this type of approach

A great deal of attention has been paid for the different requirements management approaches in CMM and agile methods. The traditional approach of having complete, consistent, precise, testable and traceable requirements may work fine in a situation where the requirements are stable but when the change rate increases, the traditional approach encounters difficult update problems [6]. The adequacy of XP requirements management practices is questioned especially when applied to component-based software development or large organizations [17]. XP is not considered to offer a generic solution that fulfills CMM requirements for requirements management [15, 16, 18]. On the other hand Turner [8] claims that CMMI and agile methods are not conflicting in requirements management process area. The results of this study confirm this latter interpretation at least in projects that fall within the scope of the assessment reported.

## 6 Conclusions

Agile software development and CMMI have been seen as conflicting views to software development. To challenge this view, this paper reported results from a study where a project using enhanced XP approach was assessed using the CMMI framework. The results of this study confirm the theoretical comparisons between XP and CMM [8, 13, 16, 21, 22] claiming that XP, by the book, does not fulfill CMMI requirements. However, the results also show that it is possible to construct a process that fulfills CMMI requirements by adding additional practices to XP. Various authors have suggested that there is a need for methods that combine agile and more traditional elements [3-10]. The results of this study confirm that it is possible to construct such methodologies in practice.

In this particular case the organization used in addition to the XP practices many traditional project management practices to initiate and steer the project and provided needed training and mentoring for the project team. The project used light documentation practices, but more documentation was done than what XP suggests. Some lightweight versions of typical traditional practices were introduced in testing and configuration management. Project team improved their processes using post mortem sessions at the end of each iteration.

This study evidenced that CMMI is one possible framework that can be used as a helping tool when building methods that combine agile and more traditional elements. Especially CMMI can be used as tool for checking that all relevant aspects are covered in the method. However, CMMI should be applied cautiously because the interpretation of the CMMI requirements is not always straightforward in the context of agile practices. One important reason for the interpretation issues was found out to be different conception of tacit knowledge in CMMI and agile methods. Despite the observed challenges, CMMI can be a valuable tool when building up processes that combine agile and traditional elements for extending the current scope of applicability of the agile methods.